# Comment on Masanari Asano et al. A model of epigenetic evolution based on theory of open quantum systems. Syst Synth Biol, 2013

*Vasily Ogryzko, Institute Gustave Roussy, Villejuif, France*
*vogryzko@gmail.com*

In a recent publication, 'A model of epigenetic evolution based on theory of open quantum systems' [1], the authors state: "We call our model quantum-like (QL) to distinguish it from really quantum models in cell biology: reducing cell's behavior to quantum particles inside a cell, e.g., Ogryzko [2, 3], McFadden and Al-Khalili [4], McFadden [5]."

*Ogryzko vs McFadden-Al-Khalili*

While I believe this statement accurately describes McFadden and Al-Khalili's position (whose main focus is on proton tunneling and base tautomerization in DNA), to characterize my own this way would be inaccurate and unfair. Regrettably, the authors appear to have been influenced by an earlier mischaracterization of my position in McFadden and Al-Khalili's paper [4]. Another author making a similar error is Melkikh [6][1]. I am concerned that this misunderstanding will be perpetuated in this nascent field for years to come.

I have never proposed to reduce cellular behavior to the quantum properties of their constituent molecules. On the contrary, my aim was always to describe the cells themselves with the operator formalism, hence the recently introduced concept of Quantum Biology at the Cellular Level (QBCL) [7]. That proposal was based on *(i)* the recognition of the limits to how much can be observed concerning an individual biological object (e.g., a single cell) and *(ii)* an analogy between self-reproduction and measurement, as noted by Wigner [8]. In its appeal to quantum theory as a more general (than classical) theory of probabilities [9], my position is, in fact, much closer to that of Asano et al than to McFadden and Al-Khalili.

My own and McFadden-Al-Khalili approach (i.e., either cell or DNA, respectively, in quantum superposition) have been directly contrasted [10]. The crucial difference between them goes to the core of what is truly new in the quantum description of reality, i.e., to the notion of quantum entanglement (or more generally, non-classical correlations).

Non-classical correlations can create unusual situations - a part B of a quantum-mechanical system A could effectively behave classically. A case in point, of course, is environmentally induced decoherence (EID), which accounts for the appearance of a classical world from an inherently quantum mechanical description via the entanglement of the system under question (e.g., B) with its environment [11, 12]. Although oversimplifying, imagine *(i)* a complex

---

[1] The text 'Goswami and Todd (1997) and Ogryzko (1997) have proposed that adaptive mutations may be generated by the environment-induced collapse of the quantum wave function that describes DNA as a superposition of mutational states. For such a mechanism to be feasible, the evolving DNA wave function must remain coherent sufficiently long for it to interact with the cell's environment' from [6], is a copy-paste from [4]

system A[2] and *(ii)* its part B[3] engaged in numerous interactions with the remaining degrees of freedom of A. The resulting reduced density matrix of B ($\rho_B$) could have its off-diagonal elements suppressed, i.e., $\rho_B$ will appear as a classical mixture.

Bottom line, *we cannot expect to fully leverage the power of quantum mechanics in biological research if we arbitrarily limit its role to describing the quantum properties of the cell's components alone - taking non-classical correlations into account means that the composite system ('the whole') can behave more quantum than its parts.* Not appreciating this point, I believe, is the principal weakness of McFadden-Al-Khalili approach [10][4]. Most of the arguments for how Life fights decoherence are moot for the same reason: namely, they focus on the (macro)molecules within cells rather than the cells themselves.

On the other hand, is it plausible to consider a living cell to be a quantum object? The lessons of decoherence theory tell us that it is not its size per se, but rather the nature of a system's coupling with the environment that matters. Fully in accord with the notions of environmentally induced decoherence and einselection [11], a starving cell weakly interacting with its environment is proposed to be in a preferred state, i.e., resistant to decoherence [7, 10]. Significantly, such preferred states will appear as a *superposition* in an alternative basis, - e.g., 'cloning basis', - which becomes important when the cell is placed in conditions allowing it to self-reproduce [10]. Ostensibly counterintuitive, this particular notion of a 'formal superposition' is distinct from the notorious Schroedinger's cat paradox [3, 7, 10], contradicting neither common sense nor observation insofar as the components of superposition (e.g., the elements of the cloning basis) are indistinguishable before the change in the environment [7, 10].

An intriguing evolutionary consequence of QBCL is 'Basis-Dependent Selection' (BDS), where *(i)* both variation and selection take place at the level of an individual organism, and *(ii)* the spectrum of heritable variations becomes dependent on the particular selection conditions [3, 7, 10, 13]. BDS is sufficiently general that it works for epigenetic variations as well [7, 10].

*Ogryzko vs Asano et al*

I would additionally like to contrast my position with that of Asano et al [1]. The authors advance the notion of *'quantum-like properties'*, as if to wash their hands of many difficulties - including complexity, multi-scale organization and decoherence - that thwart the successful application of quantum physics to biological systems. QBCL, by contrast, is based on the notion that to understand Life, *real quantum theory* - 'warts and all' - is needed, albeit applied in a far more challenging setting than usual.

In justifying my view, I would point out that quantum mechanics has been firmly established as providing a strikingly accurate description of physical reality. No experimental violations from its predictions have ever been found. Recent confirmation of the universality of quantum laws, as demonstrated by the superposition of fullerenes [14] and the detection of quantum entanglement between systems separated by km distances [15], extends the reach of quantum mechanics beyond the strictly microscopic world. The world is fundamentally quantum mechanical whereas the classical descriptions of macroscopic systems are, by contrast, only approximations.

---

[2] E.g., a living cell

[3] E.g., DNA

[4] in slightly different words, one cannot prepare nucleotide inside a living cell in a specific tautomeric form, whereas one can prepare a living cell in starved state.

Living cells contain trillions of electrons and nuclei whose physical (re)arrangements and interactions constitute their respective intracellular dynamics. As our technology continues to improve in resolution, the elucidation of intracellular dynamics starting from first (i.e., quantum mechanical) principles will become both increasingly important and attainable. An admittedly intimidating task, I assume the optimistic stance that (and not unlike the simpler cases in condensed matter physics) we can find an *approximation* that nonetheless preserves and faithfully reveal the key features of intracellular dynamics [7, 13].

I highly doubt, however, that approximations arriving at a classical description could be of much help here. While they can work well - so long as we use coarse-graining (e.g., studying a large population of individual molecular components of cells or else follow the center of mass of a cell), - when it comes to describing the intracellular dynamics of an individual cell, the scope of any coarse-graining procedure becomes severely limited. The renormalization group, widely used in condensed matter physics [16], is not applicable either, since biological systems are spatially bounded and not scale-invariant (i.e., exhibit different laws at every new level of organization [17]). The burden of proof is on those who would claim that classical approximations can supply an accurate description of intracellular dynamics without sacrificing their essential aspects. I believe, to the contrary, that any usable approximation must consider and incorporate non-classical correlations between molecular events in individual cells [7, 13].

Clearly, *real* quantum physics is needed here. The challenge, however, is to find the proper level of abstraction that would allow us to see its most important aspects without becoming lost in the details. Despite the many obstacles to creating a comprehensive and rigorous formulation of QBCL, it has already suggested some intriguing and testable consequences, such as BDS [7].

*Quantum Biology and Synthetic Biology*

The final point I would like to make will be particularly appreciated by the readers of the journal **Systems and Synthetic Biology**. In our recent paper [7], we make a case for a synergy between QBCL and Synthetic Biology. That is, by creating a minimal cell from scratch, Synthetic Biology can help to rule out alternative classical explanations for some of QBCL's predictions.

The problem with testing QBCL is that, so long as we use natural biological systems, the possibility exists that one or more classical as-yet undiscovered mechanisms could potentially account for any nontrivial prediction of QBCL[5]. Indeed, one could always argue that, whatever the nontrivial phenomenon predicted by QBCL may be, the billions of years of biological evolution could have been responsible for a classical (if cryptic) mechanism that could reproduce all of its unique features.

---

[5] e.g., BDS predicts that the biasing of stochastic events (at the cellular or molecular level) towards a more adaptive outcome is a natural property of living cells (QBCL). However, for every particular case of such phenomenon, one can alternatively suggest a specialized molecular-biological mechanism

The ultimate goal of Synthetic Biology is to rationally design and construct a biological system, under the experimenter's complete control, and in the spirit of Feynman's motto 'If I can't build it, I don't understand it'. Imagine for a moment a minimal self-reproducing cell created *from scratch* - i.e., not only its DNA, but every other molecular component was designed and synthesized de novo. Given that nothing beyond our control was involved, we can guarantee the absence of cryptic genes (or of any other mechanisms that could have emerged in evolution), alternatively accounting for the phenomenon predicted by QBCL. The observation that the minimal synthetic cell nevertheless behaves according to QBCL, would close the 'classical molecular mechanism loophole' and thus strongly support the QBCL research program.

Accordingly, it is my hope that synthetic biology will eventually serve as QBCL's natural partner, ultimately providing experimental support and validation for its principal claims.